\documentclass[%
reprint,
 amsmath,amssymb,
 aps,
prab,
]{revtex4-2}

\usepackage{graphicx}% Include figure files
\usepackage{dcolumn}% Align table columns on decimal point
\usepackage{bm}% bold math
\usepackage{algorithm,algcompatible}
\usepackage{textcomp}
\usepackage{siunitx}% Units formatting

\begin{document}

\preprint{APS/123-QED}

\title{Autonomous operation of the DIAG0 diagnostic line for 6D phase-space monitoring at LCLS-II}% Force line breaks with \\
%\thanks{A footnote to the article title}%

\author{Ryan Roussel}
\affiliation{SLAC National Accelerator Laboratory, Menlo Park, CA 94025, USA}
\author{Gopika Bhardwaj}
\affiliation{SLAC National Accelerator Laboratory, Menlo Park, CA 94025, USA}
\author{Dylan Kennedy}
\affiliation{SLAC National Accelerator Laboratory, Menlo Park, CA 94025, USA}
\author{Chris Garnier}
\affiliation{SLAC National Accelerator Laboratory, Menlo Park, CA 94025, USA}
\author{An Le}
\affiliation{SLAC National Accelerator Laboratory, Menlo Park, CA 94025, USA}
\author{William Colocho}
\affiliation{SLAC National Accelerator Laboratory, Menlo Park, CA 94025, USA}
\author{Michael Ehrlichman}
\affiliation{SLAC National Accelerator Laboratory, Menlo Park, CA 94025, USA}
\author{Yuantao Ding}
\affiliation{SLAC National Accelerator Laboratory, Menlo Park, CA 94025, USA}
\author{Feng Zhou}
\affiliation{SLAC National Accelerator Laboratory, Menlo Park, CA 94025, USA}
\author{Auralee Edelen}
\affiliation{SLAC National Accelerator Laboratory, Menlo Park, CA 94025, USA}

\date{\today}% It is always \today, today,
             %  but any date may be explicitly specified

\begin{abstract}
Characterizing the full 6-dimensional phase-space distribution of beams from the LCLS-II photoinjector is essential for understanding and optimizing downstream accelerator performance. Long-term monitoring of this distribution is equally important for detecting drifts in machine state and implementing timely corrective actions. Continuous phase space characterization during routine operation demands reliable tomographic diagnostic measurements and fast, efficient reconstruction methods. In this work, we demonstrate the first fully autonomous 6-dimensional beam-tomography system deployed on the DIAG0 parasitic beamline at LCLS-II. Using machine-learning-based control algorithms, the system autonomously configures DIAG0 and executes tomographic manipulations within operational constraints, adaptively re-optimizing beamline parameters and scan ranges in response to changes in the incoming beam. Tomographic measurements are streamed to the S3DF computing cluster where generative analysis methods reconstruct the phase-space distribution. We demonstrate that this framework produces detailed 6-dimensional beam reconstructions at a cadence of one reconstruction every 5 to 10 minutes, enabling real-time, multi-hour monitoring of injector beam evolution with unprecedented fidelity. These results represent a significant step toward fully autonomous operation of accelerator beamlines with real-time beam diagnostics for current and next-generation accelerator facilities. 

\end{abstract}

\maketitle

%\tableofcontents

\section{Introduction}
%- Detailed knowledge of the 6D beam distribution is required to effectively operate LCLS-II 
%- Continuous monitoring of the 6D distribution is needed to monitor injector health and correlate injector performance with FEL output 
%- Parasitic beamline DIAG0 available at LCLS-II, however, continuous monitoring requires autonomous operations (labor intensive) and fast, efficient phase space reconstructions (to practically handle # of reconstructions)
%- previous work: virtual diagnostics (LPS predictions, generative modeling), Bayesian optimization for one-off optimization tasks
%- overview of this work, use BO algorithms to sequentially set up and re-tune DIAG0 beamline for 6D measurements, use GPSR to predict the 6D phase space distribution

% LCLS is an FEL - which means that we need to be able to control the 6D dist. to improve and maintain performance
The Linac Coherent Light Source (LCLS) is a major user facility at SLAC that generates extremely bright X-ray radiation for scientific experiments. 
With the addition of the superconducting LCLS‑II upgrade, the facility now supports high-repetition-rate operation approaching 1 MHz, significantly increasing the average photon flux available to users. 
Because both facilities rely on free-electron lasers (FELs) to generate X-rays, their performance depends on the precise generation and control of electron beams in 6-dimensional position-momentum phase space throughout the accelerator \cite{huang_review_2007}. 
Consequently, measuring and monitoring the beam distribution at key locations along the accelerator is essential for informing online adjustments to beam generation and control processes that affect FEL performance. 
Detailed characterization of the beam structure (including beam halo) is also important for maintaining stable high-repetition-rate operation while avoiding excessive radiation from beam loss.
Furthermore, tracking the temporal evolution of the beam distribution during user operation can reveal how drifts in accelerator components influence the beam structure and whether these changes correlate with variations in FEL output.

\begin{figure*}
    \includegraphics[width=\linewidth]{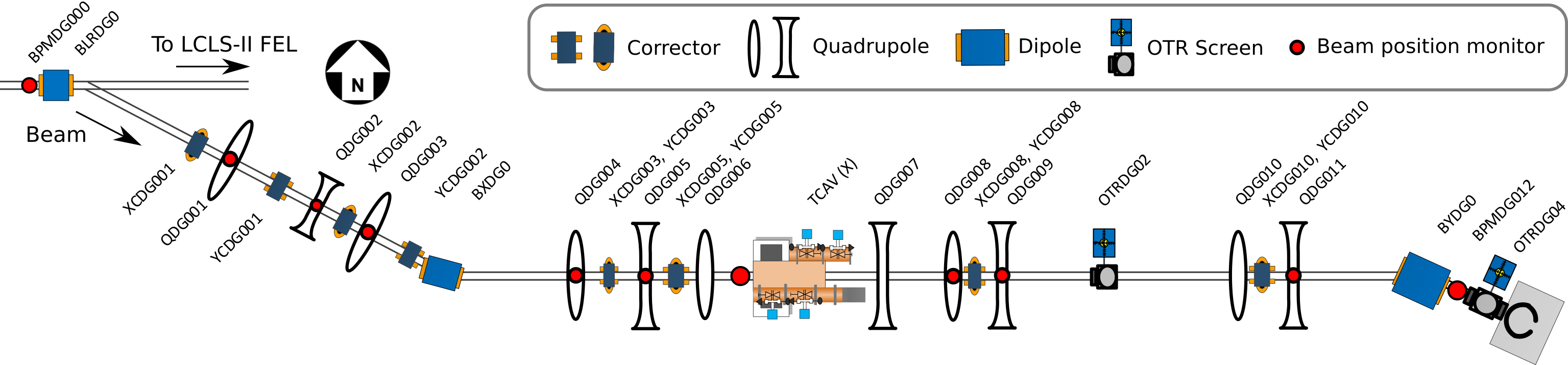}
    \caption{Schematic of the DIAG0 parasitic diagnostic line at the LCLS-II facility downstream of the LCLS-II laser heater. Beams going to this beamline are kicked transversely using an upstream fast electrostatic kicker (not pictured) and directed to the DIAG0 beamline via the magnetic septum BLRDG0.}
    \label{fig:beamline_schematic}
\end{figure*}

% we have DIAG0 to diagnose the beam distribution parasitically
To enable detailed characterization of injector beam distributions during operation, LCLS-II includes a dedicated parasitic diagnostic beamline referred to as DIAG0. 
The DIAG0 beamline transports the beam through a transverse deflecting cavity and dispersive optics that allow measurement of 6-dimensional distribution properties. However, two major challenges limit the ability to continuously monitor the phase space distribution using this beamline.

% its hard to use DIAG0 bc. of operations + hard to determine 6D dist from measurements
First, continuous operation of the diagnostic beamline requires dedicated personnel to maintain appropriate machine parameters for reliable beam transport and high-quality phase space measurements.
This need arises both when the beam distribution evolves over time due to slow drifts in operating conditions and when the accelerator is intentionally reconfigured for different operating modes.

Second, reconstructing the 6-dimensional phase space distribution using conventional tomographic methods is both experimentally and computationally expensive \cite{cathey_first_2018,hoover_analysis_2023}.
Conventional tomographic techniques and analysis methods take multiple hours to provide estimates of high dimensional distribution features, making these approaches unsuitable for online monitoring. 
Effective monitoring therefore requires analysis methods capable of producing reconstructions on time scales comparable to, or faster than, the timescale of changes in the beam distribution coming from the injector.

% in this work we demo autonomous config / measurements + do online / offline analysis of data to get 6D distribution
In this work, we demonstrate the first fully autonomous system for continuous 6-dimensional beam phase-space monitoring at LCLS-II using the DIAG0 diagnostic beamline. 
The system combines Bayesian optimization–based control algorithms \cite{roussel_bayesian_2024} to autonomously configure and maintain the beamline with Generative Phase Space Reconstruction (GPSR) \cite{roussel_phase_2023} to rapidly reconstruct the beam distribution from tomographic measurements.
Measurements are streamed to the SLAC Shared Scientific Data Facility (S3DF), where the tomographic reconstruction is performed using GPU-accelerated generative modeling techniques. 
Using this framework, the system autonomously measures and reconstructs the beam distribution during user operation with a maximum cadence of 5-10 minutes, enabling multi-hour monitoring of the beam distribution generated by the injector.

The remainder of this paper describes the autonomous control framework, the generative phase-space reconstruction method, and an experimental demonstration of continuous 6-dimensional beam monitoring using the DIAG0 beamline.

\section{Previous Work}
% BO is a good algorithm for automating DIAG0, haven't chained runs together yet
Several recent developments in accelerator control and beam diagnostics enable the autonomous monitoring framework developed in this work.

Recently, the application of Bayesian optimization (BO) algorithms \cite{roussel_bayesian_2024} has demonstrated improved convergence rates toward optimal solutions in online accelerator control contexts, thereby reducing the number of required measurements and the associated beam time needed to achieve performance objectives.
These algorithms utilize computational models called Gaussian Processes (GP) \cite{rasmussen_gaussian_2006} of objectives and constraints to inform control decisions.
Furthermore, by incorporating constraints such as beam loss limits or maximum beam size on profile monitors directly into the acquisition strategy \cite{gardner_bayesian_2014}, BO algorithms can significantly reduce the frequency of constraint violations, which is particularly important for reliable and safe autonomous accelerator operation.
BO frameworks also allow adaptive control over the trade-off between exploration, which probes uncertain regions of parameter space, and exploitation, which refines known optima. This balance can be adjusted based on the needs of a given optimization or control task.
Although Bayesian optimization has been widely used for individual tuning tasks, coordinated orchestration of multiple BO controllers to autonomously operate an entire beamline has not previously been demonstrated at SLAC.

% determining 6D dist from measurements is challenging/barrier to monitoring
Determining the high-dimensional phase space distribution of beams from experimental data has been a longstanding problem in accelerator physics.
The first 6-dimensional phase space measurement was conducted at SNS \cite{cathey_first_2018} using a set of movable slits to measure the beam distribution density at different phase space coordinates over the course of 16 hours.
Additionally, the SART \cite{andersen_simultaneous_1984} algorithm has been used to determine beam distributions from tomographic data up to 5 dimensions \cite{jaster-merz_5d_2024}, requiring approximately 1000 measurements and many hours of computing time.
As a result, it is infeasible to use these types of techniques to characterize the temporal evolution of the beam distribution during online operations.

% GPSR is demonstrated way to do this faster
More recently, the Generative Phase Space Reconstruction (GPSR) algorithm has been developed \cite{roussel_phase_2023} that combines a generative parameterization of the beam distribution with a differentiable beam dynamics model to reconstruct detailed phase space distributions from experimental data.
This algorithm has demonstrated 6-dimensional phase space reconstruction using tomographic experimental measurements via a quadrupole, dipole, and transverse deflecting cavity at the Argonne Wakefield Accelerator \cite{roussel_efficient_2024} nearly two orders of magnitude faster than previous methods.
Additional work at the Pohang-xFEL facility demonstrated 6-dimensional phase space reconstruction from tomographic measurements inside a bunch compressor \cite{kim_deployment_2025}.

% Virtual diagnostics are another possible way to do this -- but is challenging when we don't have prior data
Finally, it is important to mention virtual diagnostic methods that have been used to predict projections of the 6-dimensional phase space distribution of electron beams by combining simulation data, machine parameters, and experimental measurements \cite{emma_virtual_2021,scheinker_cdvae_2024,scheinker_conditional_2024, wolski_transverse_2022}. In these approaches, pre-trained machine learning models directly map machine settings and measured signals to predicted phase space projections, leveraging extensive training datasets assembled from prior experiments and high-fidelity, well-calibrated beam dynamics simulations.

% difference between GPSR and VD
This differs fundamentally from GPSR, which employs machine learning models to generate particle distributions in 6-dimensional phase space and relies exclusively on tomography data together with a differentiable model of the known accelerator beam dynamics to infer the underlying distribution. Rather than learning a direct input-output mapping from historical data, GPSR solves a constrained inverse problem informed by physics.

% GPSR is used in this case bc we don't have training data coverage for VD --> work does enable VD in the future
Although virtual diagnostics can provide faster predictions once trained, their accuracy depends on the availability and coverage of representative training data. GPSR is therefore particularly valuable in regimes where such datasets are unavailable or do not span the full range of beam distributions and operating conditions, which is the case for DIAG0 at present. Importantly, the autonomous measurement framework developed here establishes the foundation for systematic data collection over a wide range of operating conditions, enabling the future development of a high-fidelity virtual diagnostic.

\section{Beamline Description}
% description of DIAG0 -- ID where the reconstruction target is
An overview of the DIAG0 beamline at LCLS-II is shown in Fig.~\ref{fig:beamline_schematic}.
This beamline aims to experimentally measure the phase space distribution of the beam using quadrupoles, an S-band transverse deflecting cavity (TCAV), and a dipole spectrometer.
The transverse phase space distribution of the beam can be characterized by scanning the focusing strength of quadrupole QDG009 while measuring the beam profile at the downstream optical transition radiation (OTR) profile monitor OTRDG02.
The longitudinal profile of the beam can be measured using the TCAV, which streaks the beam in the horizontal direction while imaging the beam on OTRDG02.
By combining a horizontally streaked beam with the vertically oriented dipole spectrometer (BYDG0) the full longitudinal phase space distribution can be measured on OTRDG04.

To ensure that the transverse and longitudinal beam properties measured at OTRDG02 and OTRDG04 are representative of the beam distribution at the entrance to DIAG0, the diagnostic line is designed to have zero dispersion in both planes after the dogleg section (BLRDG0–BXDG0).
This is achieved by setting quadrupoles QDG001–QDG003 such that they produce a 180$^\circ$ phase advance in the horizontal plane before reaching BXDG0 and vertically offsetting QDG001 and QDG003 to cancel vertical dispersion.
However, achieving near-zero dispersion at the OTR screens is practically challenging because the transverse orbit through the dogleg must be closely controlled and the dispersion needs to be effectively measured.
Previous conventional phase space measurements of the beam emittance at OTRDG02 and in the main beamline directly downstream of DIAG0 have shown significant increases in projected transverse emittance due to non-zero dispersion in DIAG0.
As a result, this work aims to monitor the phase space distribution at the entrance to QDG004, downstream of the dogleg.
Future work will extend the reconstruction upstream to the entrance of DIAG0 and connect those reconstructions to phase space measurements in the main beamline. 

\section{Autonomous DIAG0 Operations}
\begin{figure*}[ht]
    \includegraphics[width=\linewidth]{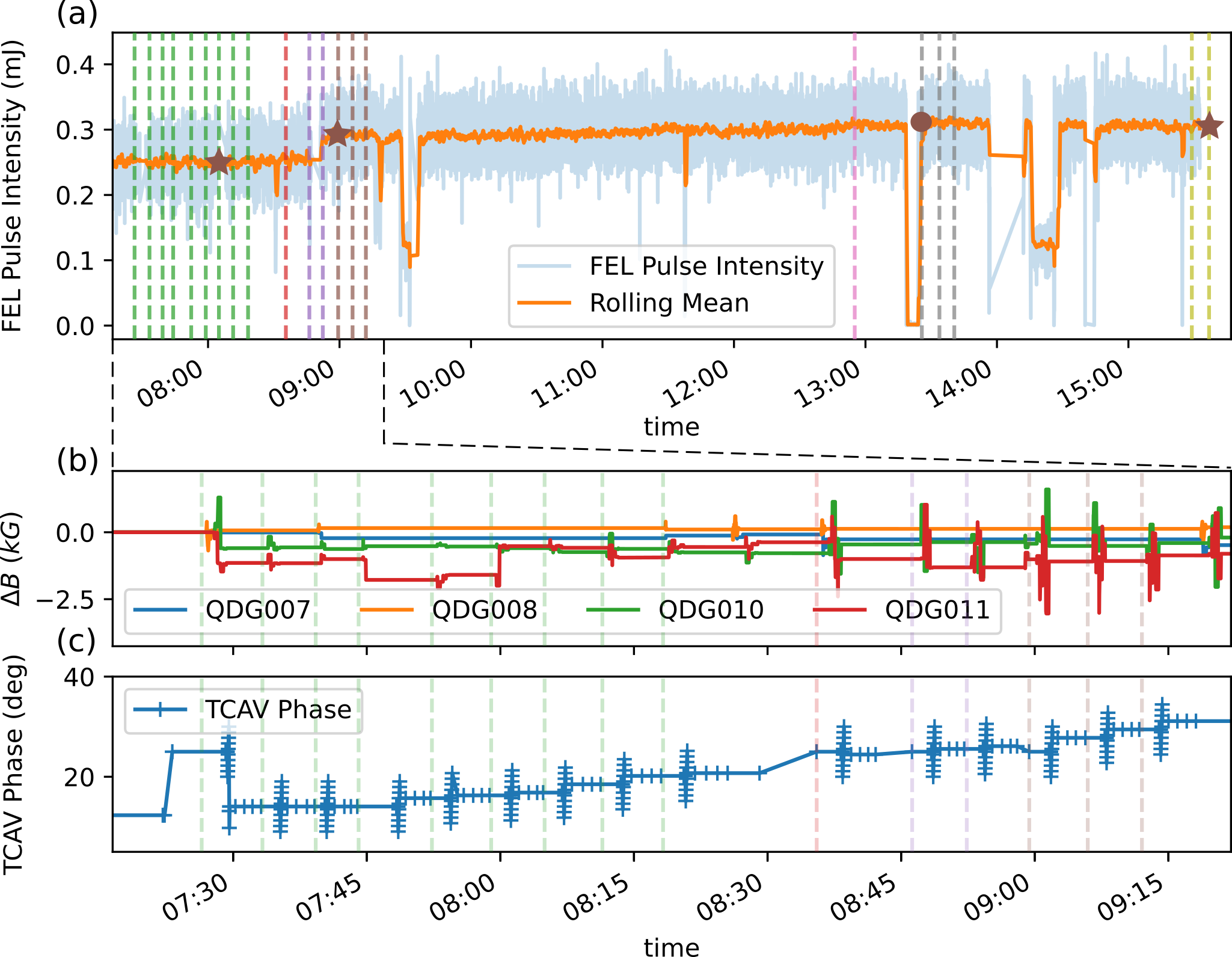}
    \caption{Time-series overview of autonomous 6-dimensional phase space measurements on the DIAG0 line during user delivery on November 25, 2025. (a) Plot of FEL pulse intensity (and rolling mean) over the course of a shift with vertical lines denoting the start of each beamline configuration and tomography workflow. Filled circle denotes measurement that serves as an example of online analysis results in Fig.~\ref{fig:online_train_pred_comparison} and Fig.~\ref{fig:online_reconstruction}. Stars denote other measurements of interest before or after time-dependent changes in the FEL pulse intensity. (b) Plot showing net change in integrated gradient over time for DIAG0 matching quadrupoles over the first 2 hours. (c) Plot showing the DIAG0 TCAV phase setpoint over the first two hours. }
    \label{fig:fel_pulse_over_time}
\end{figure*}

% this is what we did during our experimental run on 11/25
We evaluated the autonomous phase-space monitoring system during user operation on November 25, 2025 (nine-hour period). During this study, LCLS-II delivered a 70pC beam to the soft-X-ray undulator line at 35kHz for the Time-Resolved Atomic, Molecular, and Optical Science (TMO) instrument. Accelerator operation during this period prioritized stable FEL delivery, with only limited tuning of beamline parameters (outside internal steering and RF feedback loops) upstream and downstream of the DIAG0 extraction point.
The FEL pulse intensity over the course of the shift, as measured by a non-destructive gas ionization detector \cite{hau-riege_gas_2007}, is shown in Fig.~\ref{fig:fel_pulse_over_time}.
The pulse intensity was maintained at a fairly stable level throughout the shift, with a few discrete changes and interruptions causing the signal to drop out entirely or to shift.
As is typical for accelerator facilities, the average intensity exhibits slow drifts over multi-hour timescales due to factors such as temperature variations and feedback dynamics.

\begin{figure}
    \includegraphics[width=\linewidth]{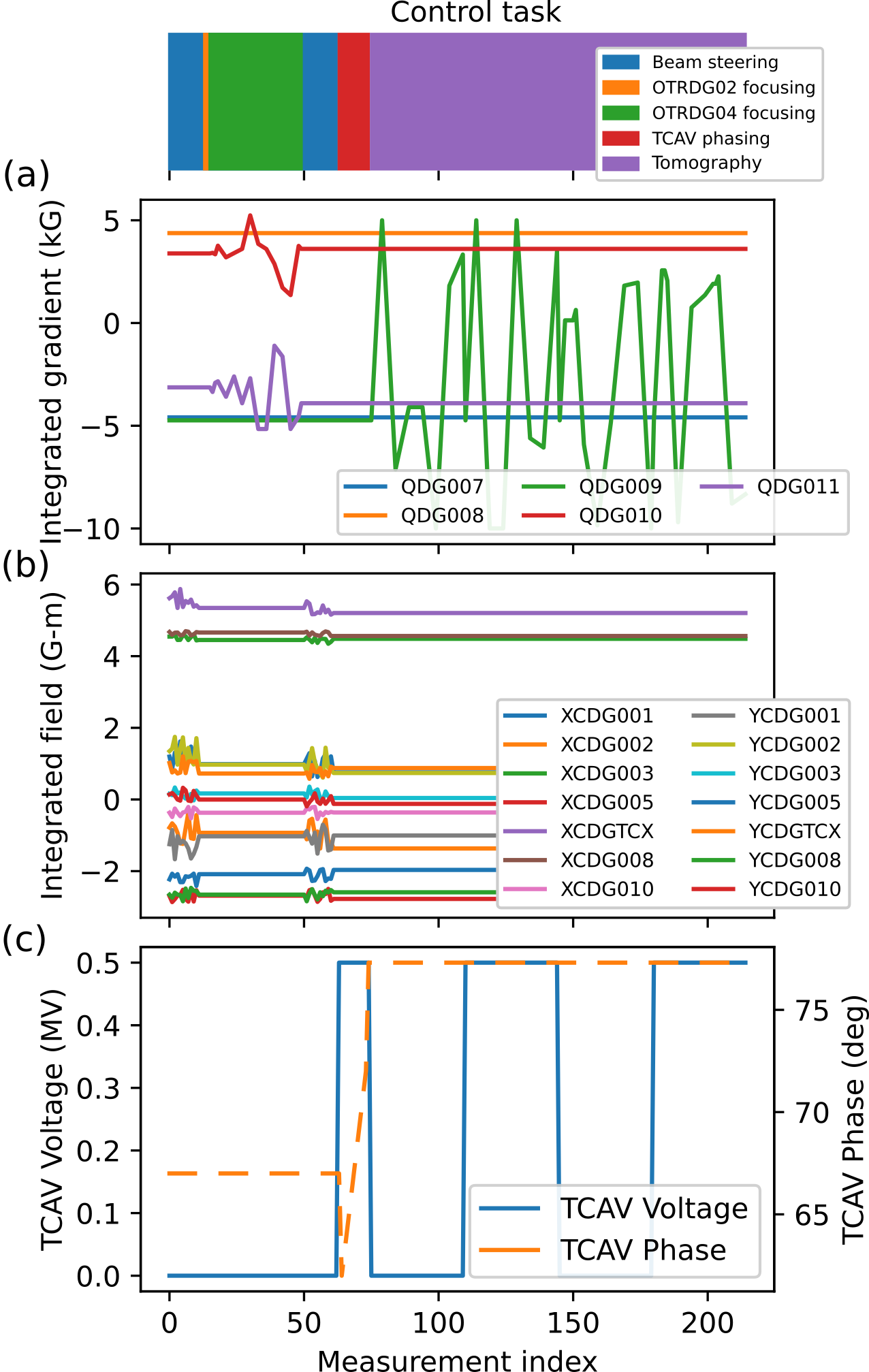}
    \caption{Evolution of beamline control parameters during a single beamline configuration and 6D phase space measurement workflow, corresponding to each control task. (a) Focusing and tomography quadrupole settings. (b) Horizontal and vertical steering magnet settings. (c) TCAV voltage and phase settings.}
    \label{fig:diag0_settings}
\end{figure}

% description of what quantifies as a "run" -- longest duration
Over the course of this shift, we conducted $7$ ``runs'' of a monitoring application that autonomously measured the phase space distribution over a period of time.
During each run, the monitor application repeated multiple beamline configuration and tomographic measurement tasks, an example of which is shown in Fig.~\ref{fig:diag0_settings}; these tasks are described in the following sections.
The monitor application was capable of responding to interruptions in beam delivery to DIAG0, pausing and restarting workflows as necessary to continue operations (see Section \ref{subsec:robustness} for details).
This enables long-term autonomous operation of the DIAG0 beamline as demonstrated in Fig.~\ref{fig:diag0_settings}.
In previous experimental shifts focused on improving robustness (see Appendix \ref{sec:appendix_auto}) we achieved a maximum autonomous operating duration of $1$ hour and $22$ minutes.

% details of run stats + justification for interruptions
In the longest of the seven runs, the monitoring application autonomously performed nine 6-dimensional phase-space measurements over 52 minutes (average cadence of 5.75 minutes per measurement).
Additional runs were used to refine the online reconstruction pipeline. Gaps between runs primarily reflect periods of workflow development and facility interruptions.
In total, we conducted twenty-one $6$-dimensional phase space measurements over the course of the shift, although in previous experiments we have demonstrated autonomous data collection of up to $42$ phase space measurements over the span of $6.6$ hours (see Appendix \ref{sec:appendix_auto}).

% tuning challenges
Autonomously operating the DIAG0 beamline such that it effectively collects high-quality tomographic information about the beam distribution requires solving multiple optimization and control problems. 
These problems include the following: (1) steering the beam through the beam pipe such that its centroid enters the centers of magnets, diagnostic screens, and the transverse deflecting cavity, (2) tuning the quadrupole magnet strengths to focus the beam at each diagnostic screen (to improve temporal and energy resolution), (3) adjusting the transverse deflecting cavity phase such that the beam arrives at the zero-crossing, and (4) conducting tomographic quadrupole scans at each diagnostic screen with the transverse deflecting cavity on and off.

% how we addressed these challenges
To address these optimization and control challenges, we employ Bayesian optimization (BO) algorithms implemented in the Python package Xopt \cite{roussel_xopt_2023}. 
BO methods are well suited to this setting: they are robust to measurement noise, can incorporate prior physics knowledge, and can be tailored to the specific requirements of each problem. 
A key operational constraint is that tuning on the DIAG0 line must not disrupt the main beamline, as this would interrupt user experiments. 
This imposes a tightly restricted parameter space, defined by limits on beam loss (and therefore radiation levels). To enforce this, we impose a constraint requiring the fractional charge loss to remain below 90\% for all control tasks.

% use of TURBO for close to optimal conditions
In addition, for optimization problems (1)–(3), we use a trust-region BO approach \cite{eriksson_scalable_2019} once the objective value approaches a specified fraction of its expected minimum. 
This adaptively narrows the search region as optimization progresses, focusing exploration on a local neighborhood around the current best solution. 
As a result, the algorithm converges more efficiently, reducing the number of iterations required to reach an optimal solution.

% starting condition for run + control task stopping conditions
Autonomous operation of DIAG0 begins with a pre-configured set of steering magnet and focusing optics parameters based on design values.
For most upstream injector configurations, this results in close, but non-optimal beam transport through the DIAG0 beamline.
Solving the individual optimization and measurement tasks is done sequentially as shown by Fig.~\ref{fig:diag0_settings}.
Each control task is stopped by either reaching a predetermined target value or a maximum number of iterations.
As a result, initial configuration of DIAG0 and tomographic measurements may take additional time as the initial configuration is outside the predetermined target values.
While successive configuration tasks may take less time as they are below the predetermined target values, in cases where disturbances to the upstream beamline or temporal drift affects the incoming beam distribution, the DIAG0 parameters also must be re-optimized to meet target objective function values.

\subsection{Beam Steering}
% beam steering problem
Beam steering optimization involves adjusting the strengths of 14 corrector magnets to center the beam at downstream beam position monitors (BPMs) that measure the horizontal and vertical position of the beam. As illustrated in Fig.~\ref{fig:beamline_schematic}, BPMs are co-located with magnetic and RF element centers, where steering requires guiding the beam through the centers of these elements while also maintaining loss-free transport through the beam pipe. 
We define the objective as minimizing the root-mean-square (RMS) beam offset in both horizontal and vertical directions, evaluated across all BPMs, as a function of the steering magnet strengths.

% custom modeling approach
To address the high dimensionality of this problem, we model each BPM signal independently as a function of the steering magnet settings. 
Specifically, we train separate Gaussian process models for each BPM, and combine samples from these models to evaluate the overall steering objective. 
This formulation leverages the approximately linear relationship between BPM offsets and steering magnet strengths, enabling the optimizer to rapidly identify irrelevant control variables (e.g., magnets located downstream of a given BPM). 
In doing so, it effectively reduces the dimensionality of the search space and improves optimization efficiency.

\begin{figure}
    \includegraphics[width=\linewidth]{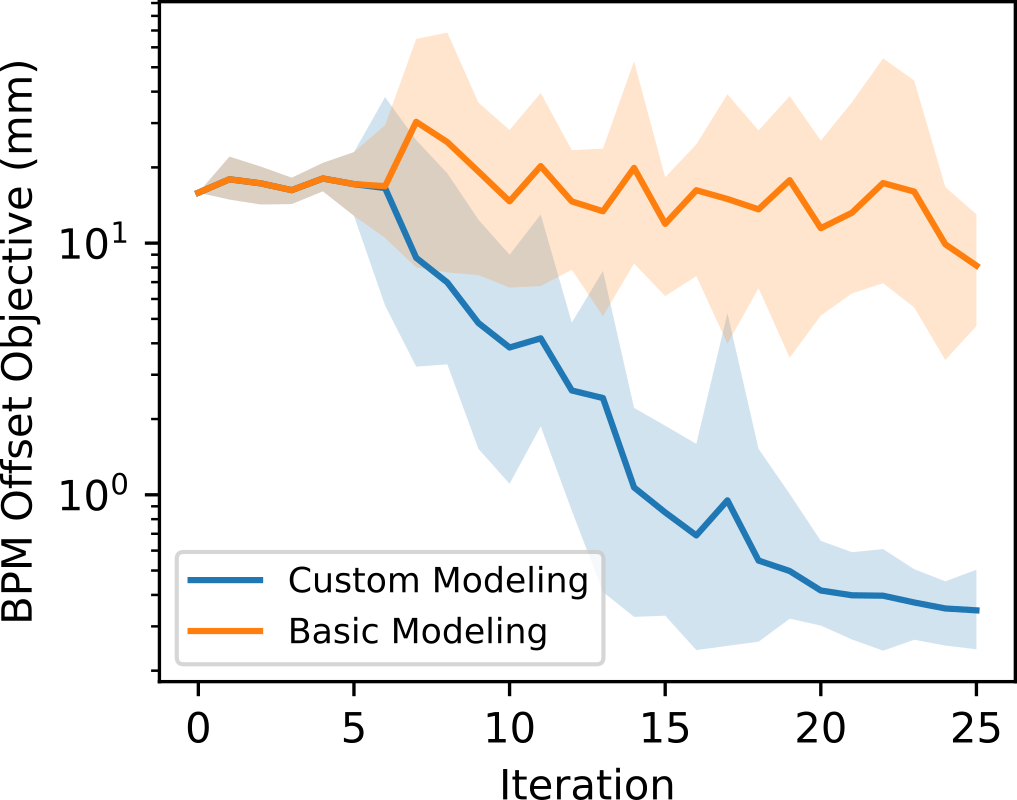}
    \caption{Bayesian optimization performance comparison between basic GP modeling of the total BPM offset objective versus independent modeling of each BPM signal in a simulated optimization of DIAG0 corrector magnets. Solid lines denote the average performance of 10 randomly initialized optimization trials while shading denotes the minimum and maximum performance at each iteration. In each trial the first 5 iterations are chosen randomly.}
    \label{fig:alignment}
\end{figure}

As shown in a simulated example in Fig.~\ref{fig:alignment}, this modeling strategy substantially reduces the number of optimization steps needed to converge to an acceptable minimum on a simulation of the DIAG0 beamline.

\subsection{Beam Focusing}
% beam focusing problem - demo of necessity
As shown in Fig.~\ref{fig:beamline_schematic}, multiple sets of quadrupoles are used to focus the beam onto the two diagnostic screens OTRDG02 and OTRDG04.
In conventional longitudinal phase space measurements, focusing the beam distribution on these screens is necessary to reduce the betatron contribution to the overall beam size when streaked due to a spectrometer or transverse deflecting cavity, thus increasing spectral or temporal resolution.
Two Bayesian optimization workflows, highlighted in Fig.~\ref{fig:diag0_settings}, utilized conventional Constrained Log Expected Improvement (logEI) \cite{ament_unexpected_2025} to focus the beam onto the two diagnostic screens.
The first workflow uses QDG007–QDG009 to focus the beam in both horizontal and vertical dimensions, via a root-mean-square (RMS) objective function, onto OTRDG02.
The second workflow uses QDG010–QDG011 to focus the beam via a total area objective function onto OTRDG04 (owing to the number of quads available and the dipole spectrometer before the diagnostic screen).
These workflows are run multiple times as needed if the final beam spot size does not reach the required target value.
The necessity of this re-tuning is exemplified in Fig.~\ref{fig:fel_pulse_over_time}(b) which shows the time-evolution of quadrupole strengths needed to maintain focus on the OTR screens.

\subsection{TCAV Phasing}
% TCAV tuning problem - demo of necessity
The aim of TCAV phasing is to adjust the phase of the RF TCAV such that the beam experiences the zero-crossing phase of the transverse kick due to the TCAV structure.
This is necessary to keep the beam on orbit when enabling and disabling the TCAV while also providing the strongest and most linear transverse kick to the beam particles as a function of time, leading to the highest temporal resolution.
TCAV phasing is commonly done by minimizing the centroid deflection between the TCAV-on and TCAV-off cases as a function of RF phase.
This process is automated via Bayesian optimization (using the same constrained logEI acquisition function as used in the focusing case), utilizing a periodic kernel function to embed the expected sinusoidal dependence of the deflection on RF phase into the GP model used to model the objective, resulting in faster convergence to the optimum.
An additional speed-up in optimization time was gained by using a brute-force, grid-based numerical optimization algorithm to maximize the acquisition function in parallel, as opposed to a sequential gradient-based algorithm.
Figure~\ref{fig:fel_pulse_over_time}(c) shows how the optimal RF phase drifted over the course of 6 hours, necessitating the re-tuning of the TCAV's RF phase during long-term operations.
Finally, data collected during this optimization is used to continuously measure the amplitude of the TCAV kick for performing phase space reconstructions.
We observed that the TCAV amplitude varied by 5-10\% for each calibration measurement, which is incorporated into each reconstruction.

\subsection{Tomographic Measurements}
% description of autonomous quadrupole scan problem
Transverse beam phase space measurements are commonly performed using tomographic techniques such as quadrupole scans \cite{mckee_phase_1995, yakimenko_electron_2003}. In these methods, the focusing strength of a quadrupole magnet is varied while recording the horizontal and vertical beam distributions on a downstream screen, and the resulting projections are used to reconstruct the transverse beam matrix or phase space distribution.

In practice, quadrupole scan setpoints are typically chosen manually by operators, often by defining a grid of values over a selected range of quadrupole strengths. To accurately estimate the beam matrix parameters, quadrupole strengths must span a sufficiently wide range to rotate the phase space while also capturing points around the minimum beam size (where the phase advances most rapidly). At the same time, the settings must avoid beam losses and keep the beam distribution within the bounds of the diagnostic screen. Because the horizontal and vertical beam dynamics often differ, identifying a range of quadrupole strengths that satisfies these conditions for both planes can be challenging. Additionally, changes in the incoming beam distribution due to machine drift or the application of tomographic manipulations (i.e., enabling or disabling a transverse deflecting cavity) further change the ideal range of quadrupole strengths. As a result, operators frequently need to repeat the measurement multiple times to adjust the scan range before obtaining suitable data for reconstruction.

Achieving autonomous emittance measurements therefore requires a more systematic method for selecting quadrupole settings that can account for these constraints and adapt to variations in the incoming beam distribution, all while ensuring that measurements effectively gather enough information about the phase space distribution.

\begin{figure}
    \includegraphics[width=\linewidth]{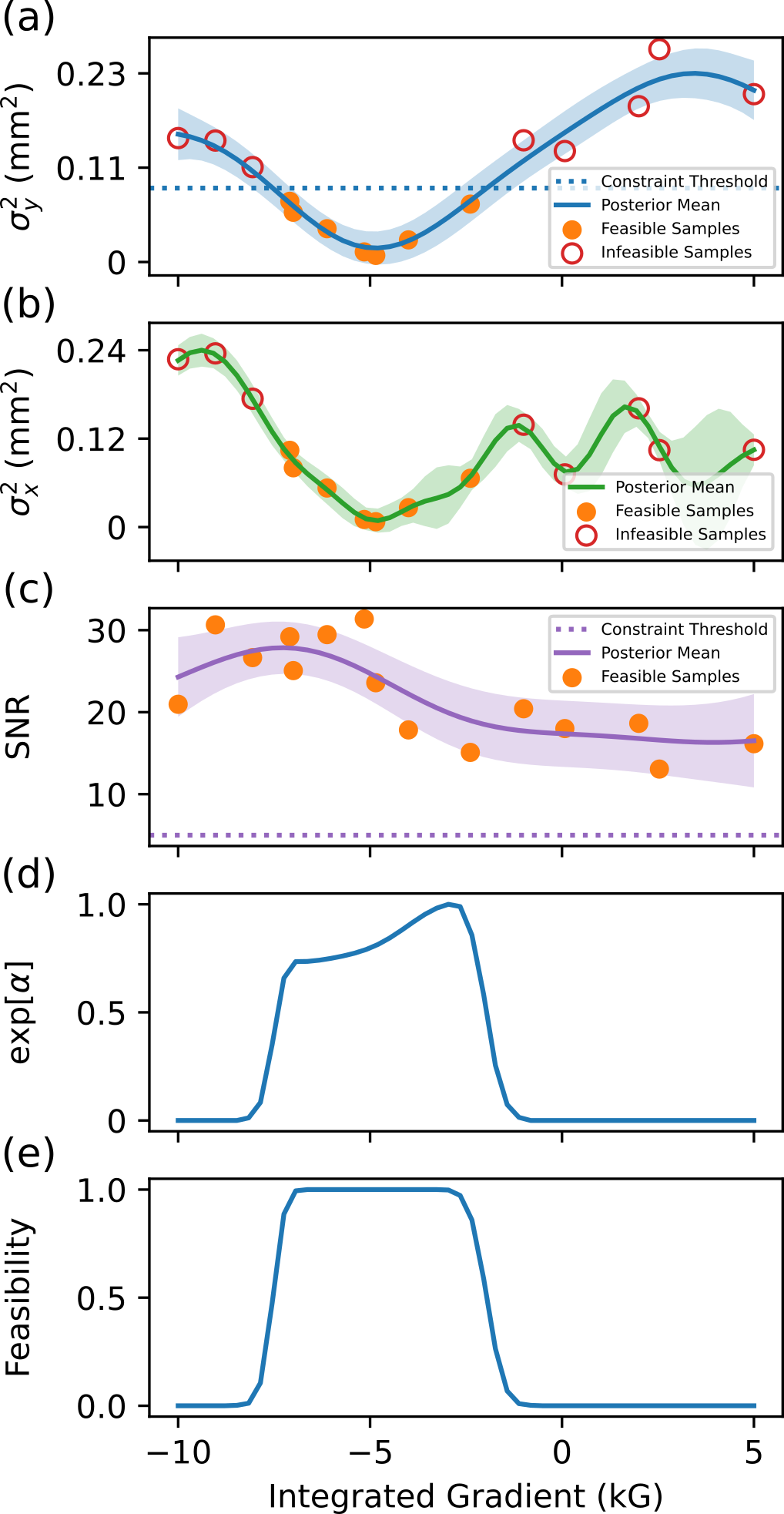}
    \caption{Visualization of autonomous emittance algorithm quadrupole strength selection process midway through the autonomous quadrupole scan measurements. (a) Gaussian process (GP) model of vertical beam sizes squared as a function of quadrupole integrated gradient with maximum beam size constraint shown. (b) GP model of horizontal beam sizes squared (not used when targeting vertical phase space measurements). (c) GP model of signal-to-noise (SNR) ratio of the beam distribution on the screen. (d) Acquisition function which is maximized to select the next quadrupole setting. (e) Predicted probability of satisfying constraints.}
    \label{fig:automatic_emittance}
\end{figure}

% autonomous BO solution (general)
To achieve this, we utilize Bayesian optimization with a constrained Upper Confidence Bound (UCB) \cite{brochu_tutorial_2010} acquisition function to iteratively determine the quadrupole settings used to perform beam tomography targeting the horizontal and vertical phase spaces (see \cite{roussel_demonstration_2023} for details).
The UCB acquisition function allows the user to explicitly control the algorithm's preference for finding an optimum vs. characterizing the input parameter space.
By using a UCB acquisition function that aims to minimize the beam size with a higher preference for exploration ($\beta=10^4$), the BO algorithm will choose quadrupole strengths that both identify the minimum beam size and sample around the minimum as seen in Fig.~\ref{fig:automatic_emittance}.

To avoid making measurements where the beam is too large to properly measure the RMS profile we include two observational constraints into the acquisition function.
These include a maximum RMS beam size constraint, proportional to the smallest beam size observed, and a minimum signal-to-noise ratio (SNR) constraint, which measures the ratio of peak beam intensity on the screen to the magnitude of the background noise.
We incorporate these constraints into the acquisition function by creating independent Gaussian process models of each constraint function and then weighting the UCB acquisition function by the probability that all constraints are satisfied (referred to as the feasibility, shown in Fig.~\ref{fig:automatic_emittance}(d)).

%The addition of observational constraints like maximum beam size and minimum beam losses lowers how frequently the BO algorithm chooses quadrupole strengths that violate these conditions.

% autonomous BO solution (practical)
We start the autonomous measurement with a coarse, evenly-spaced scan over a subset of the possible focusing strength range to get a rough characterization of the beam size as a function of focusing strength.
Using this initial dataset we then run BO with UCB targeting the horizontal beam size for a set number of iterations.
Finally, we run a fixed number of iterations targeting the vertical beam size, leveraging both the initial coarse scan data and data gathered during the horizontal BO run to improve sample efficiency.
In addition to making beam size measurements at the quadrupole strengths determined by BO, we make beam size measurements at a set number of intermediate measurement points between each BO-determined set point, allowing us to quickly gather more data without additional BO steps (which can be relatively slow).
During experimental runs, a single autonomous quadrupole scan takes approximately $30$–$45$ seconds to conduct $30$ profile measurements, a subset of which are used to inform the measurement of the transverse phase space.

% expanding autonomous quad scans to 6D tomography
As demonstrated in previous applications of GPSR \cite{roussel_efficient_2024, kim_deployment_2025}, combining the data measured from quadrupole scans at non-dispersive and dispersive diagnostic screens with TCAV off/on states can be sufficient to fully reconstruct the 6-dimensional phase space distribution.
In order to perform $6$-dimensional tomographic measurements using the DIAG0 beamline, we repeat autonomous quadrupole scans using QDG009 for both OTRDG02 (non-dispersive) and OTRDG04 (dispersive) diagnostic screens, once with the transverse deflecting cavity off and on, resulting in $4$ quadrupole scan datasets that are used to reconstruct the distribution, as shown in Fig.~\ref{fig:diag0_settings}.
This leverages the capabilities of the autonomous quadrupole scan to autonomously determine optimal scan points as both the TCAV state and the location of the diagnostic screens affect the correct quadrupole strengths for effective tomography.

\subsection{Robustness, Failure Recovery}
\label{subsec:robustness}
%- machine state monitoring
%- different recovery conditions
%- tenacity package
Autonomous operation of an entire beamline over a long period of time requires multiple layers of machine state monitoring and error handling to prevent and recover from interruptions to beamline operations.
To handle these issues, we utilized the \textsc{tenacity} package, which allows the programmer to customize whether and how procedures are retried or not when specified errors are raised, including specifying the number of retry attempts, the time interval between attempts, and the messaging to the user during each retry.
Depending on the type and duration of the interruption, different recovery strategies were used to continue the autonomous monitoring workflow without human intervention.

% short period interruptions
The first class of interruptions were short-duration interruptions due to machine status or analysis errors.
For example, machine protection systems may interrupt beam delivery due to radiation detection exceeding predefined limits, or beam profile fitting analysis workflows fail to converge to an acceptable optimum.
These errors are stochastic in nature, and are often quickly cleared by operators to resume beam delivery or can be solved by re-running analysis.
In these cases, individual measurements or workflow processes are retried at a constant frequency over the span of a few minutes in an attempt to restart a particular task.

% long term interruptions
If errors persist for longer than this period, the entire beamline configuration and tomographic measurement process is repeated from the initial beamline settings (recorded before each configuration and measurement run).
This is necessary since we assume that the beam distribution entering DIAG0 is static over the course of the tomographic measurement, thus collecting data for more than a few minutes can be influenced by drift in the machine over longer time periods.
This assumption also requires restarting the entire beamline configuration and measurement workflow when changes are made to beamline parameters upstream of the DIAG0 line, which is avoided by performing this demonstration during a period of user beam delivery where beamline parameters upstream of DIAG0 are rarely altered outside of minor perturbations.

\section{Reconstruction Process and Results}
%- workflow overview, data pre-processing, transfer and processing on S3DF
%- Single measurement results
%- Long term tracking results w/ UQ
In this section, we describe the reconstruction process used during experimental measurements. We present both online and offline reconstruction results, as well as post-run results using experimental data. 

\subsection{Generative Phase Space Reconstruction}
% GPSR overview
GPSR utilizes a neural-network-based parameterization of the beam distribution in 6-dimensional phase space and a beam dynamics model of the DIAG0 beamline and diagnostics to solve for the beam distribution at the entrance of DIAG0.
In this work, we used the Cheetah \cite{kaiser_bridging_2024} simulation package, which implements the backwards-differentiable particle tracking necessary for reconstructing the beam distribution using GPSR.
To ensure tracking accuracy, particle tracking in Cheetah was benchmarked to Bmad \cite{sagan_bmad_2006} beam dynamics tracking commonly used for accelerator modeling at SLAC.
The beamline parameters and calibrations corresponding to tomography measurements are converted into simulation units and then applied as element settings in the Cheetah beam dynamics simulation.

%GPSR specifics
Arbitrary beam distributions are parameterized by using a densely connected neural network (4 layers of 100 neurons each, LeakyReLU function scaled by 0.25) to transform normally distributed random numbers into real particle coordinates in phase space.
This architecture is used to generate macro-particle distributions containing 100k particles each, balancing computational speed with reconstruction fidelity.
These particle distributions are tracked through the Cheetah beam dynamics simulation of the DIAG0 beamline onto simulated versions of the OTRDG02 and OTRDG04 diagnostic screens.

\subsection{Online Reconstructions}
% 6d_data_1764090357
% describe reconstruction workflow
%- show example of reconstruction comparison to experimental measurements + reconstructed distribution
%- describe mean fitting of all images
%- incl. comparison with raw image to demonstrate robustness of fitting (?)
%- "This enables detailed tracking of the beam evolution through the beamline, for example, we can take the distribution ..." show / discuss beam tracking results from online reconstruction
We developed an automated pipeline for real-time reconstruction of beam distributions by utilizing the SLAC Shared Science Data Facility (S3DF), an on-site high-performance computing cluster.
After each phase space tomography measurement, beamline parameters (quadrupole, dipole, steering magnet currents, TCAV phase, and amplitude) and diagnostic measurements (OTRDG02, OTRDG04 images) are pre-processed and converted into a PyTorch serialized dataset file that is placed in a staging directory (see Appendix \ref{sec:appendix_gpsr} for pre-processing details).

On the S3DF side, an analysis job submitted via SLURM continuously monitors a designated directory on S3DF for new datasets.
When a new dataset file is detected, the GPSR reconstruction workflow is triggered along with logging of processed and failed attempts to analyze a given file to ensure robust operation. 
This job used a single S3DF GPU node (NVIDIA A100, 80~GB RAM).
Each GPSR workflow took approximately 6 minutes to determine the phase space distribution.

\begin{figure*}
    \includegraphics[width=\linewidth]{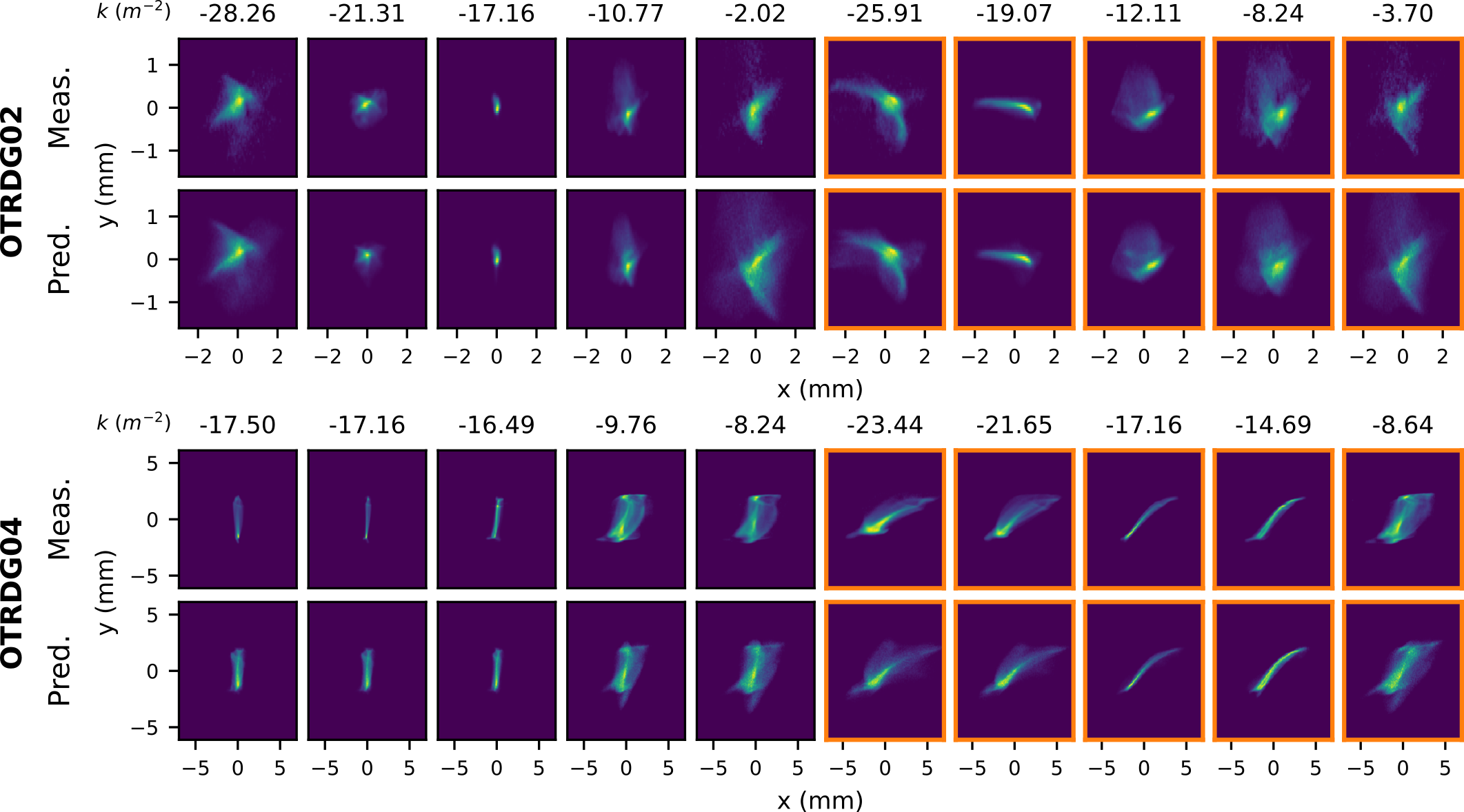} % 1764105944
    \caption{Comparison between measurements (top row) and online GPSR predictions (bottom row) of the transverse beam profile at OTRDG02 and OTRDG04 during tomography scans starting at 13:25 PST. Each column denotes a different quadrupole strength used during autonomous measurements at each OTR screen with the TCAV off and on (red borders). Note that the TCAV streaks the beam in the horizontal plane, while the dipole bends the beam in the vertical direction for OTRDG04.}
    \label{fig:online_train_pred_comparison}
\end{figure*}

\begin{figure*}
    \includegraphics[width=\linewidth]{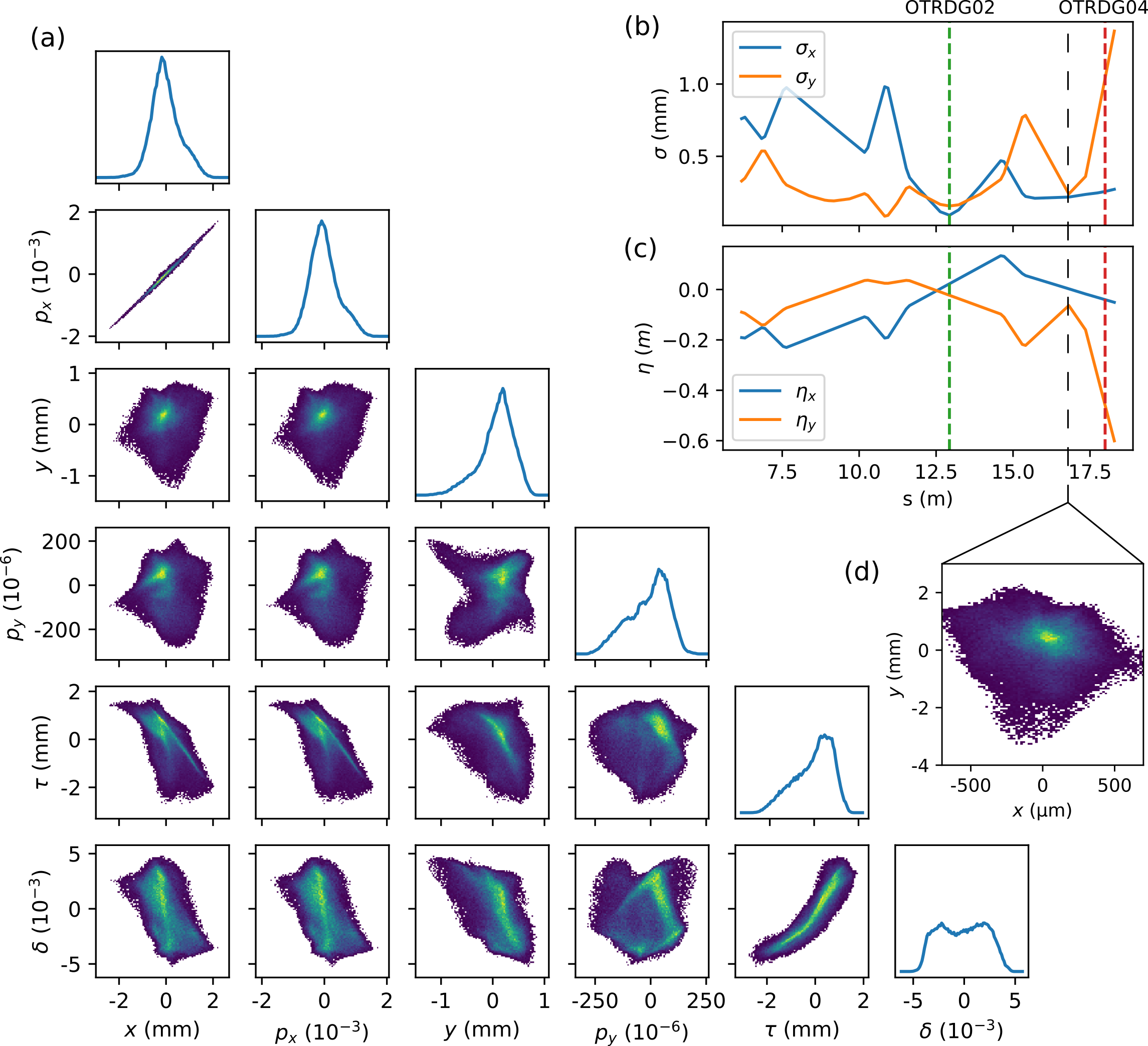} % 1764105944
    \caption{(a) Projections of reconstructed beam distribution at the entrance to QDG004 along the principal phase space coordinates. Once reconstructed, the distribution can be tracked through a lattice using a beam dynamics simulation that predicts RMS quantities such as the transverse beam size (b) and the horizontal dispersion (c). A detailed beam distribution can also be predicted at locations of interest along the DIAG0 beamline (d).}
    \label{fig:online_reconstruction}
\end{figure*}

An example of results obtained from a single phase space measurement using online GPSR is shown in Fig.~\ref{fig:online_train_pred_comparison} and Fig.~\ref{fig:online_reconstruction}.

Figure \ref{fig:online_train_pred_comparison} shows a comparison between post-processed beam profile measurements at OTRDG02 and OTRDG04 and predicted beam profiles from the GPSR process.
Predicted beam profiles are generated by propagating the reconstructed beam distribution at QDG004 through the DIAG0 Cheetah simulation to the simulated diagnostic screens.
In the example shown here, we see good qualitative agreement between the processed experimental measurements and the reconstruction predictions.
In some cases, lower density features predicted by GPSR are not in agreement with the experimental measurements.
This may be due to several reasons, first, for cases where the beam size is the largest the signal-to-noise ratio of pixel intensities on the screen may be small such that portions of the beam distribution are removed during background removal.
Second, the reconstruction aims to determine an average beam distribution that matches all images simultaneously, meaning that shot-to-shot fluctuations in the fine features of the beam distribution may be averaged out in the reconstruction, which causes discrepancies between measurements and predictions.
Even with these considerations, agreement between the predictions and images shows predictive capabilities beyond conventional RMS reconstructions.

Figure \ref{fig:online_reconstruction} shows projections of the 6-dimensional phase space distribution reconstructed at QDG004 along the principal phase space position-momentum coordinates.
The GPSR reconstruction algorithm produces a macro-particle distribution identical to those used in particle tracking simulations, allowing arbitrary visualization of distributions projected along any set of 2- or 3-dimensional axes.
Additionally, this particle distribution can be tracked through a downstream beamline, allowing prediction of RMS phase space quantities such as beam size and dispersion, Fig.~\ref{fig:online_reconstruction}(b) and (c) respectively, as well as detailed predictions of phase space distributions at points of interest along the accelerator, as shown in Fig.~\ref{fig:online_reconstruction}(d).
This type of online reconstruction can be used to perform optimization of upstream beamline elements, such as the photoinjector and laser heater parameters, or used to predict beam evolution through the rest of the main beamline down to the FEL undulator sections.
For these applications, it is possible to reduce the fidelity of the reconstruction (fewer particles tracked, fewer images, or lower resolution images) to reduce the computational cost of phase space measurements for faster predictions.

\subsection{Offline Reconstruction Results}
%- after the shift we reprocessed / analyzed the data and examined the evolution of 6d phase space distribution over time
%- UQ discussion
%- show / discuss evolution of y-yp, y-p, tau-p phase spaces
%- show discuss RMS quantity values over time
%- discuss possible correlations with FEL pulse intensity
After the shift, we conducted offline, higher-quality reconstructions of the phase space distributions to examine how the distribution evolved over the course of the experimental run.
This analysis included slight changes to image post-processing as well as minor changes to the neural network structure used in GPSR (see Appendix~\ref{sec:appendix_gpsr} for a discussion).

The largest difference between offline and online analysis results presented here is the use of ensembling methods to estimate uncertainties in the reconstructed distribution.
To estimate the uncertainty in the reconstruction, multiple, independent runs of the GPSR analysis algorithm are executed on the same experimental dataset.
This method utilizes different random initial starting points in parameter space and the inherent stochastic nature of gradient optimization to reach multiple potential solutions that solve the reconstruction problem.
It should be stressed that this method is a heuristic, and provides only a rough estimate of the relative uncertainty in the reconstruction due to information content and measurement noise.
This approach captures algorithmic uncertainty associated with the non-convex reconstruction problem but does not account for systematic uncertainties in diagnostics or shot-to-shot beam fluctuations.
A more robust treatment of uncertainty would account for shot-to-shot variations in the beam distribution and systematic uncertainties in diagnostics and accelerator control elements.
In this work, ensembling took place in parallel on multiple GPUs to create ensembles of 10 reconstructions for each experimental dataset.

\begin{figure}
    \includegraphics[width=\linewidth]{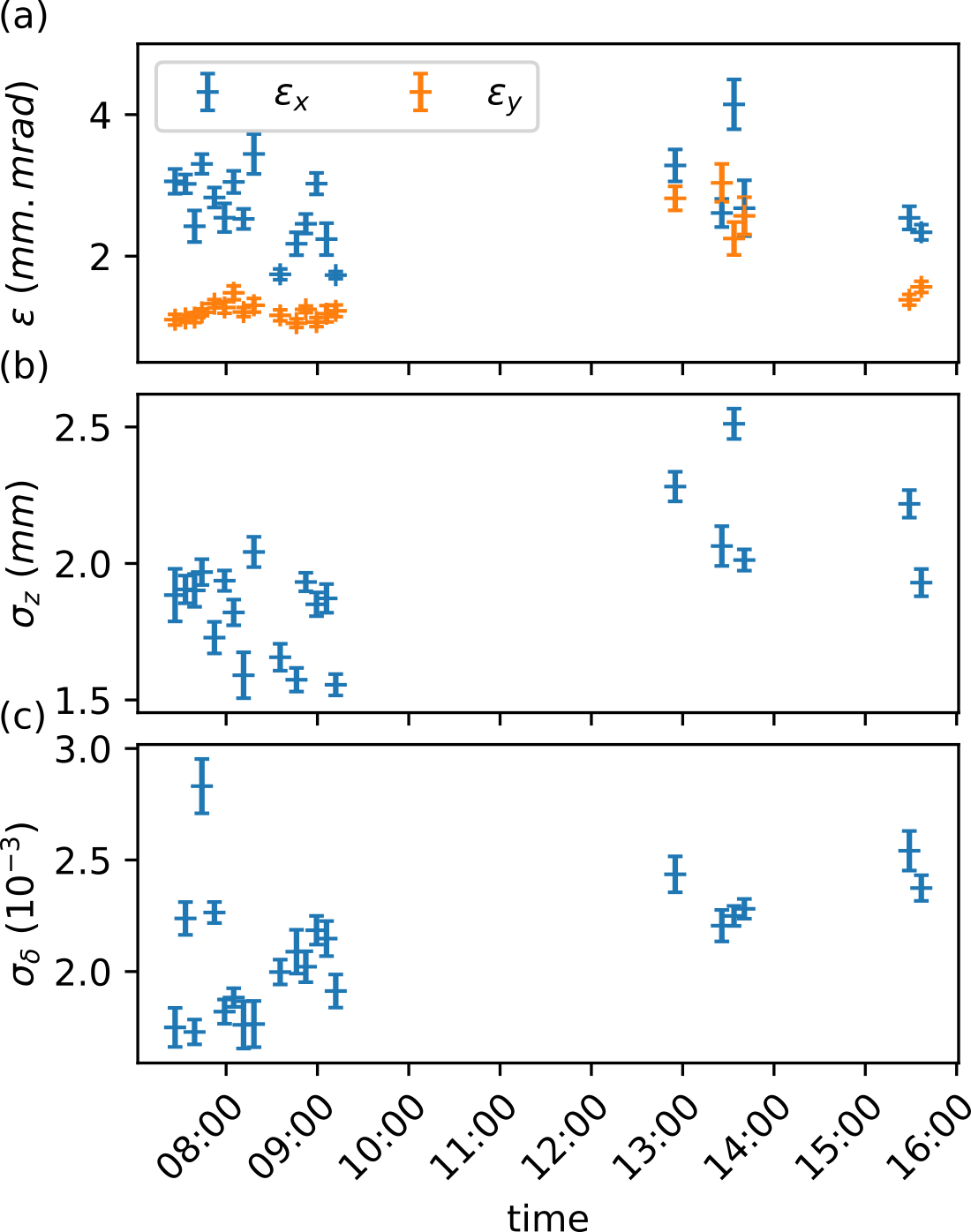}
    \caption{Visualization of scalar measurements of the beam distribution at QDG004 over time including core beam emittance (a), bunch length (b), and energy spread (c). Points include 1$\sigma$ confidence intervals provided by uncertainty quantification estimates. Core of the beam is determined by selecting 90\% of the macroparticles that are closest to the origin in normalized 6-dimensional phase space.}
    \label{fig:rms_statistics_over_time}
\end{figure}

Figure \ref{fig:rms_statistics_over_time} shows the evolution of scalar RMS quantities of the beam distribution over time.
We observe that the RMS beam emittances are relatively stable in the y-direction for the first two hours but then grow during the second half of the shift, while the measured horizontal emittance is relatively noisy.
Similar trends are evident for the bunch length and energy spread.

\begin{figure*}
    \includegraphics[width=\linewidth]{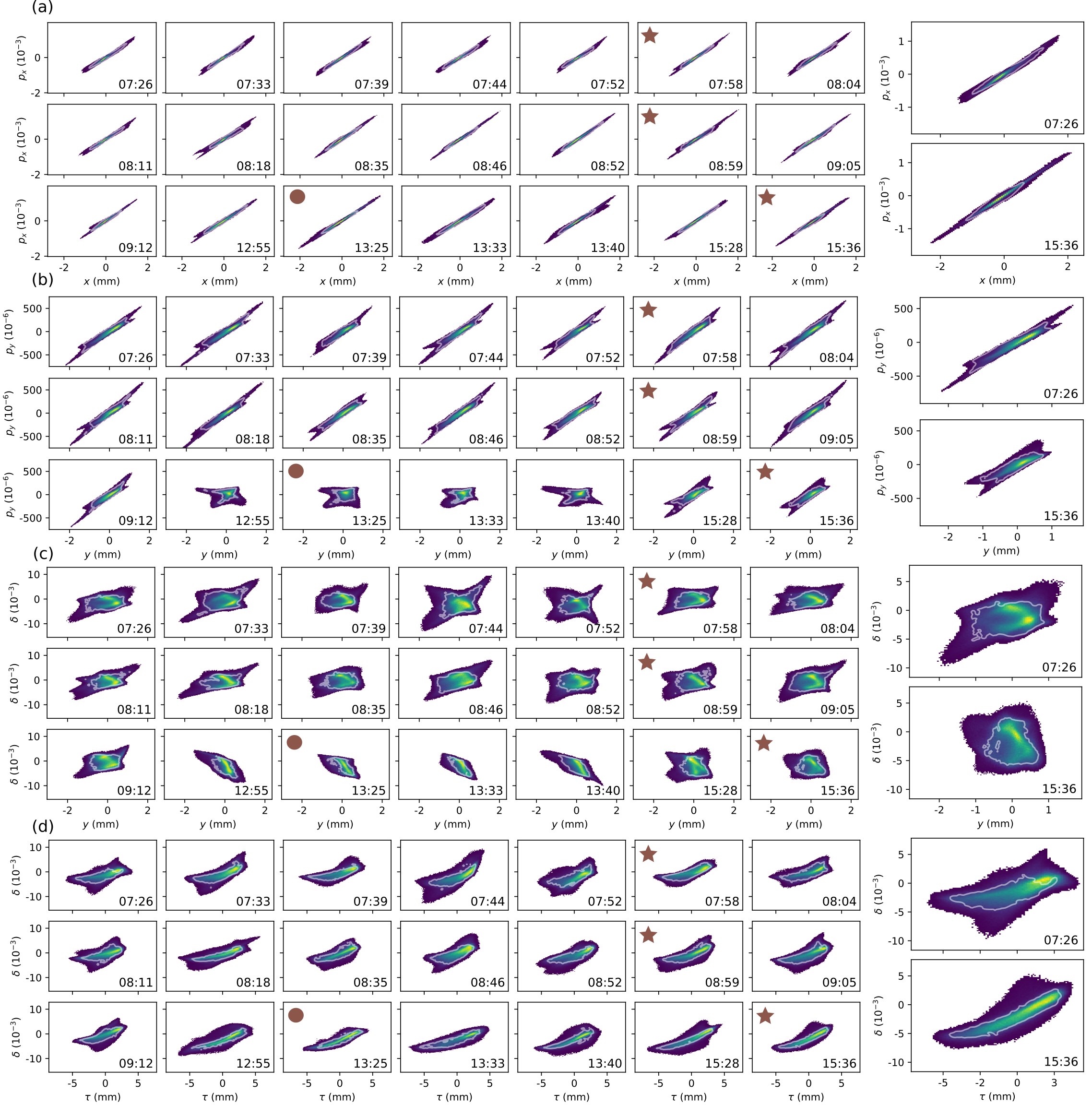}
    \caption{Evolution of selected 2-dimensional phase space distributions over time (lower left, PT). (a) Horizontal phase space distribution, (b) vertical phase space distribution, (c) vertical-energy deviation distribution, and (d) longitudinal phase space distribution. Rightmost plots denote phase space distributions at the beginning and end of the shift. Stars and filled circles denote reconstruction results highlighted in Fig.~\ref{fig:fel_pulse_over_time}, with experimental data shown in Fig.~\ref{fig:online_train_pred_comparison} and Appendix~\ref{sec:appendix_additional_comparisons}. Contours denote region above confidence factor level greater than 2.}
    \label{fig:beam_dist_over_time}
\end{figure*}

The phase space reconstruction allows us to also visualize how detailed features of the 6-dimensional beam distribution evolve over time.
Figure \ref{fig:beam_dist_over_time} shows the time-evolution of a subset of 2-dimensional projected phase spaces over the course of the shift.
Note that while these 2-dimensional projections are along the phase space coordinate axes, arbitrary projections of the distribution are possible because the reconstruction produces a macroparticle distribution.

Here we observe similar behavior as is described in Fig.~\ref{fig:rms_statistics_over_time} but with more detail.
Prior to 12:55 PST the overall vertical phase space distribution remains largely unchanged, with minor differences in the core beam distribution.
The vertical phase space distribution then changes again between 13:40 and 15:28, returning back to the initial distribution, albeit with a less correlated core distribution and a slight rotation.
On the other hand, there seems to be significant changes in the structure of the $y-\delta$ distribution which goes from a seemingly two bunch structure where multiple cores are observed at different energies to a smoother, well correlated distribution with non-linear curvature.

Finally, while the overall structure of the longitudinal phase space distribution remains roughly constant, the rotation and strength of the nonlinear correlation changes over time.
It should be noted that changes to features in the reconstructed distribution are commensurate with changes in the experimental measurements, which can be viewed in Appendix \ref{sec:appendix_additional_comparisons}.

In Fig.~\ref{fig:beam_dist_over_time}, we estimate the confidence in the predicted phase space density using a statistical confidence metric based on the ensemble of reconstruction results.
We collect the predicted density of particles at a given point in phase space from each ensemble result into a distribution and measure the quantiles of that distribution to compute a normalized confidence metric $C = q_{50} / (q_{90} - q_{10})$ where $q_x$ denotes the $x^{th}$ quantile of the probability distribution.
We consider a confidence metric of greater than 2 to exhibit high confidence in the predicted density (meaning that the uncertainty in the beam density at that location is 2x smaller than the predicted density).
Regions of phase space that meet this criterion are surrounded by contours in Fig.~\ref{fig:beam_dist_over_time}.

We observe that there is significant confidence in the reconstruction results near the core of the beam distribution.
On the other hand, our reconstruction shows significantly less confidence in the reconstructed beam halo.
This may be explained by a lower signal-to-noise ratio for the beam halo in the experimental images compared to the beam distribution's core density.

\section{Discussion}
\label{sec:discussion}
% - improvements to autonomous control via RL / prior means for BO
% - anomaly detection
% - LLM integration for autonomous operations
\subsection{Autonomous operations}

This work represents a preliminary demonstration of autonomous operations, but challenges remain for true 24/7 continuous operation during user and machine development shifts.
For example, tomographic measurements assume a near-constant incoming beam distribution, which requires upstream accelerator parameters to be constant throughout the duration of the tomographic measurements.
Identifying when measurements are possible, and restarting measurements when significant upstream changes are present is necessary to ensure that the constant incoming beam distribution assumption is valid.

Additionally, detecting and responding to a variety of anomalous beam behavior is necessary for maintaining continuous operation over a long period of time.
In this work we identified and responded to basic interruptions and issues encountered during a small number of shifts.
Increasing the coverage of autonomous operations requires handling increasingly complex issues that need to be addressed.
This could potentially be addressed in the future via anomaly detection workflows and may be a use case for artificial intelligence-based agents for making high-level operational decisions (i.e., pausing acquisition, running optimization workflows, running analysis workflows, etc.), such as Osprey \cite{hellert_agentic_2026}.

Finally, while BO algorithms were used in this initial work due to their flexibility, robustness to noise, and ease of implementation, alternative algorithms could be used to improve convergence speed and robustness.
In particular, reinforcement learning algorithms could be easily trained to address some of the tuning tasks, leveraging the same differentiable beam dynamics model used for performing GPSR.
Additionally, the model could be used as an informative prior inside the context of BO to reduce the frequency of constraint violations and improve convergence as done in \cite{boltz2025leveraging}.
These improvements would result in faster configuration of DIAG0 and tomographic measurements, increasing the temporal resolution of time-dependent distribution monitoring.

% discuss comparison between emittances measured here and previous measurements
\subsection{Reconstruction results}
We observe that the RMS beam attributes obtained from the DIAG0 reconstruction (Fig.~\ref{fig:rms_statistics_over_time}) differ from previously reported beam measurements from the LCLS-II injector measured at the main beamline \cite{zhou_lcls-ii_2026}.
We think this is partially due to the DIAG0 optics dispersion issue, which causes emittance growth, as observed with previous conventional measurements of the beam emittance at screens in DIAG0. Further DIAG0 optics development will be performed separately, with the goal making a consistent emittance measurement between the main beamline and the DIAG0 line.

It is also instructive to examine the systematic differences between using simplified moment-based representation of beam distributions and high-fidelity reconstruction techniques. As discussed in Appendix \ref{sec:gaussian_comparison}, reconstructions based on simplified multivariate normal assumptions about the beam distribution can yield significantly different emittance estimates compared to those incorporating additional distributional detail.

These considerations motivate future studies that directly compare simultaneous phase space measurements at DIAG0 and along the main beamline.
Future work will simultaneously measure the phase space distributions in DIAG0 and at screens directly downstream of the DIAG0 septum magnet.
Reconstructions of the beam distribution from DIAG0 will include BPM signal data, allowing us to account for dispersion-based emittance growth in DIAG0 due to non-ideal beam orbit in the dogleg section.
By propagating the reconstructed beam distribution to diagnostics in the main beamline and comparing predictions of the beam profile at the diagnostics to experimental measurements, we can validate or calibrate the DIAG0 beam dynamics simulations to these joint measurements in multiple beamlines.

\section{Conclusion}
In this work, we have described the first demonstration of autonomous beamline operation and monitoring of the 6-dimensional phase space distribution of beams after the laser heater portion of LCLS-II.
We have shown that Bayesian optimization algorithms can be tailored towards solving a variety of unique optimization and control problems and can be orchestrated to autonomously operate a moderately sized beamline without human intervention on the hour timescale.
By coupling this with the GPSR algorithm on S3DF HPC, we are able to produce real-time estimations of the phase space distribution at a cadence of one reconstruction every 5-10 minutes.
We demonstrated that this workflow can be used to identify precise changes in the incoming phase space distribution during user beam delivery, which in the future could be correlated with changes to upstream beamline parameters and the downstream LCLS-II FEL pulse energy.
With further improvements to beam dynamics modeling and sustained autonomous operation of DIAG0, this work provides a path toward continuous characterization of long-term accelerator performance under user operating conditions.
Additionally, this capability represents an important step toward autonomous accelerator diagnostics and real-time characterization of high-dimensional beam distributions in next-generation accelerator facilities.

\begin{acknowledgments}
Conceptualization, R.R., and A.L.E.; Data curation, R.R., G.B., and C.G.; Formal analysis, R.R., and C.G.; Funding acquisition, A.L.E., and Y.D.; Investigation, R.R., G.B., C.G., D.K., W.C., M. E., and Y.D.; Methodology, R.R. and D.K.; Software, R.R., G.B., C.G., and D.K; Experiment, R.R., G.B., C.G., and D.K.; Supervision, A.L.E.; Validation, R.R., G.B., C.G., and D.K; Visualization, R.R., G.B., C.G., and D.K; Writing---original draft, R.R., G.B., C.G., and D.K.; Writing---review and editing, R.R., G.B., C.G., D.K., W.C., M.E., Y.D., and A.E. All authors have read and agreed to the published version of the manuscript.

This work is supported by the U.S. Department of Energy, Office of Science under Contract No. DE-AC02-76SF00515.

\end{acknowledgments}

\appendix

\section{Long-term Autonomous Phase Space Monitoring}
\label{sec:appendix_auto}
\begin{figure*}
    \includegraphics[width=\linewidth]{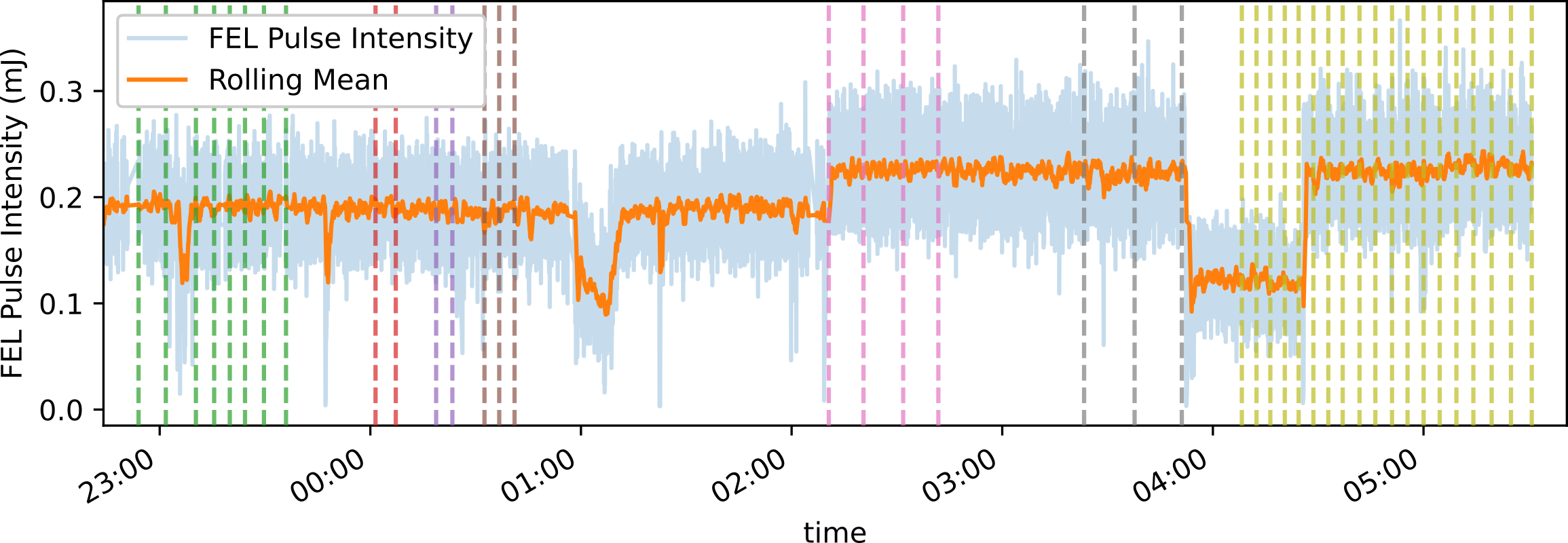}
    \caption{Time-series overview of autonomous 6-dimensional phase space measurements on the DIAG0 line on November 22, 2025. Plot of FEL pulse intensity (and rolling mean) over the course of a shift with vertical lines denoting the start of each beamline configuration and tomography workflow.}
    \label{fig:11_22_fel_pulse_over_time}
\end{figure*}

Prior to the study described in the main text, we performed additional tests to maximize the duration of uninterrupted autonomous operation on DIAG0.
Figure~\ref{fig:11_22_fel_pulse_over_time} shows a similar plot to Fig.~\ref{fig:fel_pulse_over_time} where vertical lines denote the beginning of DIAG0 configuration and 6-dimensional phase space measurement tasks. As seen here, the final run measured the phase space distribution without human input over 1 hour and 22 minutes.
Overall, during the shift we conducted 42 phase space measurements over a period of 6.6 hours.
\section{GPSR Workflow Details}
\label{sec:appendix_gpsr}
Here we describe post-processing and analysis details of the GPSR workflow.

After collecting the raw images from OTRDG02 and OTRDG04 several steps are taken to pre-process the data for use in GPSR.
Each image is processed according to the following steps: (1) Background subtraction (using beam-off images), (2) Median filtering, (3) Thresholding (using the triangle-threshold method), (4) image cropping and centering, (5) image pooling to reduce image size.

After pre-processing a subset of images gathered during the autonomous emittance measurement process are used in reconstructing the beam distribution.
These images are selected by first estimating the transverse beam size in each axis, then selecting the images with the smallest beam size plus a fixed number of images on either side of that minimum while excluding images that had a peak signal-to-noise ratio above a preset value.
This collection of images is then evenly sub-sampled again to ensure a fixed number of final images for each quadrupole scan.
The resulting dataset contained 5 images for each quadrupole scan, where sets of images for each screen had the same image sizes.

For offline GPSR, we expanded the size of the neural network from 4 layers of 100 neurons each to 4 layers of 300 neurons each such that the complexity of predicted beam distributions was increased at the cost of additional training cost.
Empirically, this did not seem to have a significant qualitative effect on GPSR's ability to match experimentally measured distributions.

\section{Selected Prediction Comparisons}
\label{sec:appendix_additional_comparisons}
\begin{figure*}
    \includegraphics[width=0.8\linewidth]{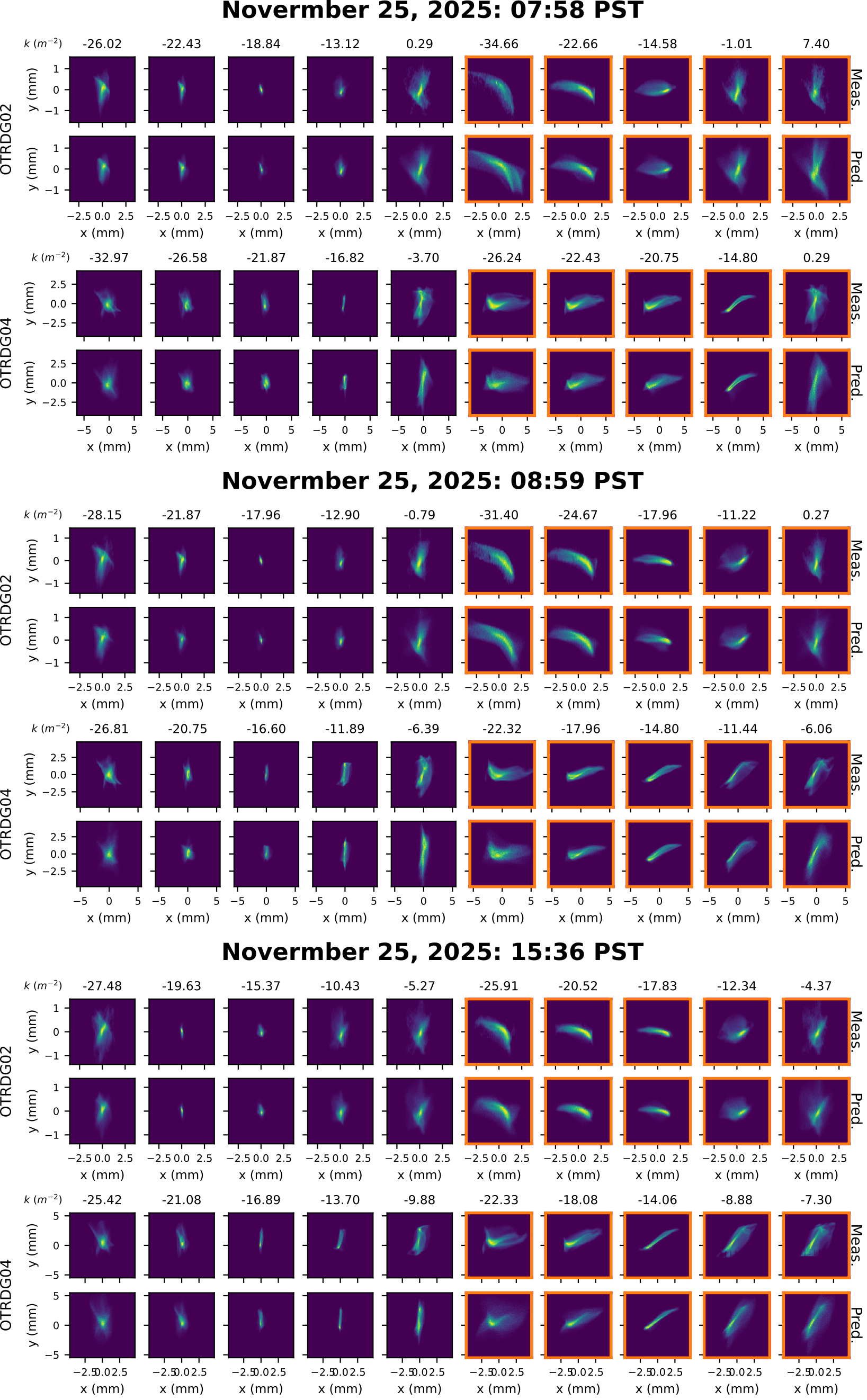} % 1764105944
    \caption{Comparison between measurements (top row) and online GPSR predictions (bottom row) of the transverse beam profile at OTRDG02 and OTRDG04 during tomography scans. Each column denotes a different quadrupole strength used during autonomous measurements at each OTR screen with the TCAV off and on (orange borders). }
    \label{fig:train_pred_comp_selected}
\end{figure*}

Figures~\ref{fig:train_pred_comp_selected} show comparisons between offline reconstruction predictions and experimental data gathered at selected times identified by stars in Fig.~\ref{fig:fel_pulse_over_time} and Fig.~\ref{fig:beam_dist_over_time}.

\section{Comparing GPSR and Multivariate Normal Distribution Fitting}
\label{sec:gaussian_comparison}
In this section, we compare the results of using conventional multivariate normal (MVN) descriptions of the beam distribution to flexible, generative representations when fitting experimental measurements.
We can constrain GPSR to generate only MVN distributions and fit these to the same experimental dataset used for the full phase space reconstruction. This can be achieved by setting all of the activation functions in the transformer neural network to be linear, which reduces distribution generation to MVNs.

\begin{figure*}
    \includegraphics[width=0.8\linewidth]{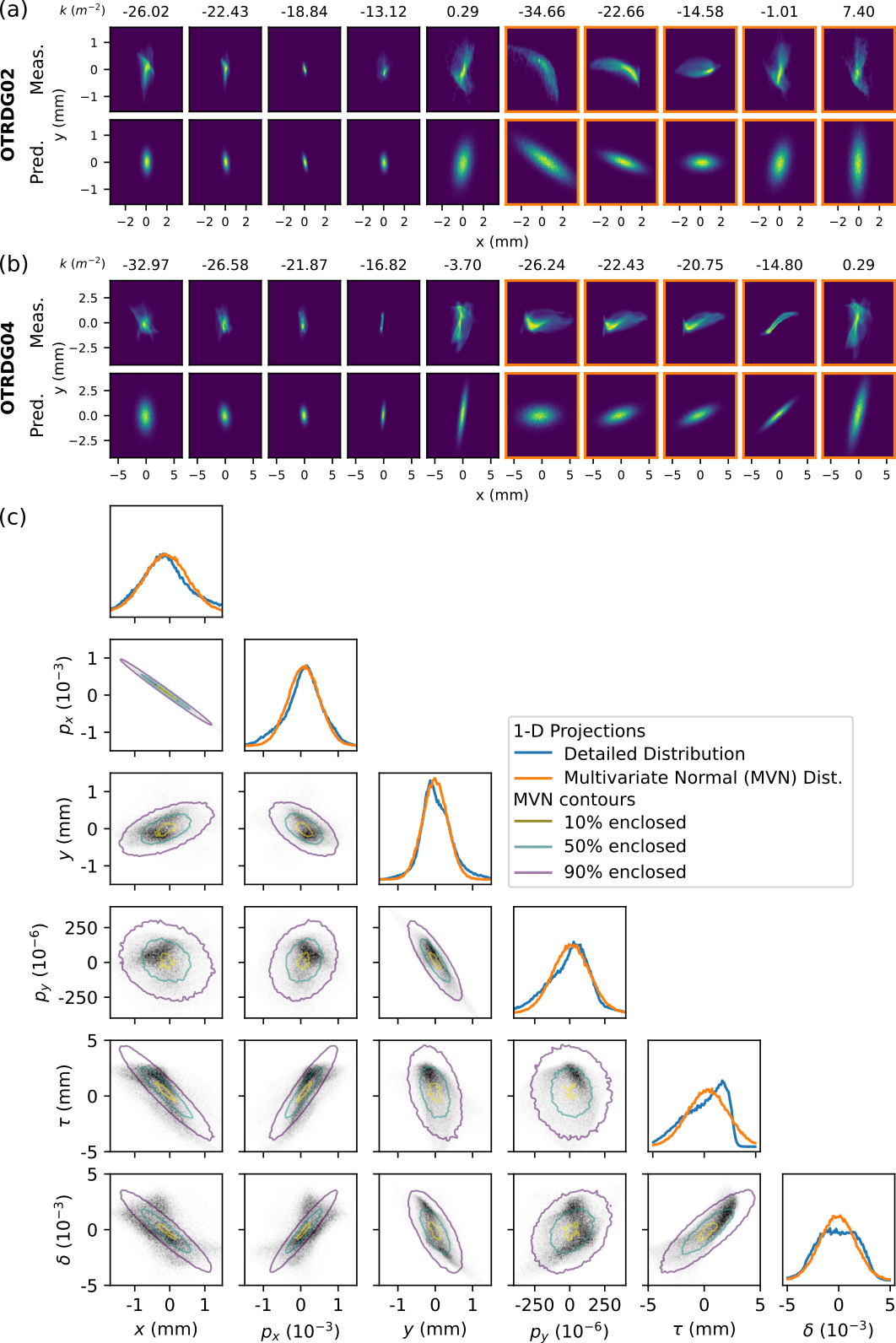} % 1764105944
    \caption{Comparison between measurements and Multivariate normal predictions of the transverse beam profile at OTRDG02 (a) and OTRDG04 (b) during tomography scans. Each column denotes a different quadrupole strength used during autonomous measurements at each OTR screen with the TCAV off and on (orange borders). (c) Comparison between projections of the reconstructed distribution using detailed GPSR (grey colormap) and Multivariate normal (contours) beam distribution.}
    \label{fig:offline_reconstruction_gaussian_comparison_paper}
\end{figure*}

Figure~\ref{fig:offline_reconstruction_gaussian_comparison_paper} shows the result of fitting an MVN model to a representative set of tomography scans and compares it to detailed, generative fitting of the same data. 
While the MVN reconstruction reproduces the overall extent of the beam, it fails to capture the detailed structure observed in the measured images (Fig.~\ref{fig:offline_reconstruction_gaussian_comparison_paper}(a,b)). In particular, fine-scale features and correlations present in the data are not represented within the Gaussian approximation.

This limitation becomes more evident when comparing the MVN distribution to the detailed reconstructed phase space distributions. As shown in Fig.~\ref{fig:offline_reconstruction_gaussian_comparison_paper}(c), the one-dimensional projections of the MVN and full GPSR reconstructions appear similar, indicating that marginal distributions alone are insufficient to distinguish between models. However, the two-dimensional structure—where correlations and non-Gaussian features reside—is not accurately reproduced by the MVN model.

These differences lead to substantial discrepancies in derived beam parameters. The second-order moments, computed from the macroparticle ensembles, differ significantly between the two reconstructions. For example, the horizontal and vertical normalized emittances (excluding dispersive contributions) are $0.5 / 1.4$ mm·mrad for the MVN reconstruction, compared to $2.8 / 1.6$ mm·mrad for the full GPSR reconstruction.

This comparison highlights that agreement in projected profiles does not guarantee accurate recovery of underlying phase space structure when computing emittances. Incorporating higher-order features through more expressive generative models is therefore essential for reliable inference of beam properties from experimental data.

% The \nocite command causes all entries in a bibliography to be printed out
% whether or not they are actually referenced in the text. This is appropriate
% for the sample file to show the different styles of references, but authors
% most likely will not want to use it.
%\nocite{*}

\bibliography{diag0_autonomous_6d_reconstruction}% Produces the bibliography via BibTeX.

\end{document}